\documentclass[aps,prl,longbibliography,twocolumn,superscriptaddress,showpacs]{revtex4-1}

\usepackage{CJK}

\usepackage{graphicx}

\usepackage{amsmath}
\usepackage{dcolumn}

\usepackage{bm}
\usepackage[normalem]{ulem}
\usepackage{color}
\usepackage{hyperref}

\usepackage{epstopdf}
\newcommand{\bew}{\begin{widetext}}
\newcommand{\ew}{\end{widetext}}

\newcommand{\ii}{{\rm i}}
\newcommand{\bx}{\mathbf{x}}

\newcommand{\bq}{\mathbf{q}}
\newcommand{\bv}{\mathbf{v}}

\newcommand{\br}{\mathbf{r}}

\newcommand{\bff}{\mathbf{f}}
\newcommand{\bu}{\mathbf{u}}

\newcommand{\bbr}{\mathbf{r}}

\newcommand{\beq}{\begin{equation}}
\newcommand{\eeq}{\end{equation}}
\newcommand{\beqn}{\begin{eqnarray}}
\newcommand{\eeqn}{\end{eqnarray}}
\newcommand{\pp}{\partial}
\newcommand{\dd}{{\rm d}}

\newcommand{\cO}{{\cal O}}

\newcommand{\cG}{{\cal G}}

\newcommand{\la}{\langle}

\newcommand{\ra}{\rangle}

\newcommand{\vnab}{{\bf \nabla}}

\begin{document}
\title{Packed swarms on dirt: two-dimensional incompressible flocks with quenched and annealed disorder}
\author{Leiming Chen}
\email{leiming@cumt.edu.cn}
\address{School of Material Science and Physics, China University of Mining and Technology, Xuzhou Jiangsu, 221116, P. R. China}
\author{Chiu Fan Lee}
\email{c.lee@imperial.ac.uk}
\address{Department of Bioengineering, Imperial College London, South Kensington Campus, London SW7 2AZ, U.K.}
\author{Ananyo Maitra}
\email{nyomaitra07@gmail.com}
\address{Laboratoire de Physique Th\'eorique et Mod\'elisation, CNRS UMR 8089,
	CY Cergy Paris Universit\'e, F-95032 Cergy-Pontoise Cedex, France}
\author{John Toner}
\email{jjt@uoregon.edu}
\affiliation{Department of Physics and Institute of Theoretical
Science, University of Oregon, Eugene, OR $97403^1$}
\affiliation{Max Planck Institute for the Physics of Complex Systems, N\"othnitzer Str. 38, 01187 Dresden, Germany}
\date{\today}

	\begin{abstract}
	{We show that incompressible polar active fluids can exhibit an ordered,  coherently moving phase even in the presence of quenched
disorder in two dimensions. Unlike such active fluids with {\it annealed} (i.e., time-dependent) disorder {\it only}, which behave like equilibrium ferromagnets with long-range interactions, this robustness against quenched disorder is a fundamentally non-equilibrium phenomenon. The ordered state   belongs to a new universality class, whose scaling laws we calculate using three different renormalization group schemes, which all give scaling exponents within 0.02 of each other, indicating that our results are quite
accurate. Our predictions can be quantitatively tested in readily available artificial  active systems, and  imply that biological systems such as cell layers can move coherently \emph{in vivo}, where disorder is inevitable.}
	\end{abstract}
	
\maketitle

 One of the most significant discoveries in statistical mechanics in the $20$th century was the ``Mermin-Wagner-Hohenberg" theorem \cite{MW}, which says that continuous symmetries cannot be spontaneously broken in equilibrium two-dimension{a}l systems at finite temperature. This implies, in particular, that magnetic  systems with continuous rotational symmetry like the XY or Heisenberg models cannot exhibit long-range polar order characterized by a non-zero magnetization in two dimensions  (2D).

It is precisely for this reason that  it was so surprising to learn that 2D {\it active} systems {\it can} spontaneously develop long-range polar order even in the presence of noise \cite{vicsek_prl95, toner_prl95, toner_pre98, toner_pre12, Chate_DADAM_rev}. In particular,  polar self-propelled particles  moving over a frictional substrate (a system often described as a ``dry polar active fluid") can ``flock"; that is, form a state with a non-zero average velocity $\langle\bv\rangle$, even when perturbed by noise.

In equilibrium systems, it is known that {\it quenched} disorder \cite{Harris, Geoff, Aharonyrandom,Dfisher} -- that is, disorder that is time-{\it independent} -- is even more destructive of long-range order than thermal noise, which is, of course, time-{\it dependent}. Indeed, even arbitrarily small quenched disorder destroys long-ranged ferromagnetic and crystalline order in all spatial dimensions $d\le4$ \cite{Harris, Geoff, Aharonyrandom,Dfisher}.

The above discussion raises the obvious question: can {\it active} systems retain long-range order even in the presence of {\it quenched} disorder \cite{toner_prl18, toner_pre18,Duan21, Dor21, Ro21, Peruani1, Peruani2, Peruani3, Bartolo2021, Bartolo1}? It has been shown \cite{toner_prl18, toner_pre18} that for three-dimensional dry polar active systems with quenched disorder, long-range polar order (i.e., a non-zero average velocity $\langle\bv\rangle$) can survive such quenched disorder. But in 2D, only quasi-long-range polar order (i.e.,  $\langle\bv\rangle$ vanishes as a power of system size $L$ as $L\to\infty$)
was found \cite{toner_prl18, toner_pre18, peruani}.

 In this paper, we report that it is possible to achieve true {\it long-range} polar order in 2D
 in dry polar active systems with quenched
disorder, if those systems are {\it incompressible}.

 There are many ways to experimentally realize incompressibility in active systems. One is to make them very dense
so that the effective compressibility of the flockers vanishes  \cite{CLT_Ncomm, Ano_pol}.
 An even more realizable incompressible system is a  suspension of swimmers in a fluid confined in a narrow channel.
 Here, the active agents ``inherit" the incompressibility of the background fluid \cite{BrottoLauga, Bricard, Ano_pol}. Strictly speaking,  swimmer suspensions differ  from the systems we consider here, due to the presence of an extra hydrodynamic variable in the former: the density of the swimmers. However, if the swimmers in the channel  are constantly  being born and dying \cite{Toner_Malthus} or can switch between an active, motile state and a passive, immotile one,
so that the number of {active} swimmers is
not globally conserved, the dynamics of the system {\it is} described by the theory presented here.

Finally, spin-systems interacting via dipolar interactions have the same long-time, large-distance
properties as magnets with a  divergence-free
constraint on the magnetization \cite{Kashuba, CLT_Ncomm}. It has further been shown \cite{Ano_asym, CLT_malthus1, CLT_malthus2} that the hydrodynamics of spin-systems with nonequilibrium and non-reciprocal asymmetric interactions are described by the equations originally constructed for (Malthusian) flocks \cite{Toner_Malthus}. Therefore, the hydrodynamics of dipolar magnets with asymmetric exchange interactions \cite{Ano_asym} is equivalent to incompressible flocks \cite{CLT_Ncomm, CLT_incom2}.  If such a magnetic system {were} subject to quenched disorder, it would again be described by the theory we develop here.

Motivated by all of the examples above, we consider here
incompressible flocks in 2D
\cite{CLT_Ncomm} on a substrate with quenched disorder.  We also include the effects of
``annealed" disorder (i.e., time-dependent noise), which proves to be subdominant for the equal-time correlations, but gives time-dependence to the fluctuations, which would otherwise be
static. Considering the full nonlinear theory, we demonstrate that the moving phase of such
a flock can sustain long-range polar order. A  similar result was found,
 based on
a purely {\it linearized}  theory, for number-conserving active suspensions  with quenched
disorder \cite{Ano_disord} (which, as noted  above,
differ somewhat from our system).

Using a dynamic renormalization group  (DRG) analysis,
we calculate the long-time, large-distance scaling of the fluctuations  $\bu( \bbr,t)$ of the local active fluid velocity $ \bv(\bbr,t)$ about its mean value $\la \bv \ra\equiv v_0 \hat{\bf x}$, where we've defined our coordinate system so that ${\bf \hat{x}}$ is along the mean velocity spontaneously chosen by the system.
 This analysis leads to
 scaling laws for the correlations of $\bu(\bbr,t) \equiv \bv(\bbr,t)-v_0 \hat{\bf x}$ which we obtain using  three different  DRG
schemes: two different $d=(d_c-\epsilon)$-expansions,
which we term ``hard continuation'' and ``soft continuation'',  and an uncontrolled
expansion in exactly
2D.

Specifically, we show that the $u$-$u$ correlations are given by
\beq
\langle \bu(\br,t)\cdot\bu(\mathbf{0},0)\rangle=|y|^{2\chi}\cG_{_{Q}}\left({|x|\over |y|^\zeta}\right)
+|y|^{2\chi'}\cG_{_{A}}\left({|x|\over |y|^{\zeta}},{|t|\over |y|^{z}}\right)\, ,\label{Correl1}
\eeq
where
$\cG_{_{A,Q}}$ are scaling functions that are universal up to an overall multiplicative factor, corresponding to the annealed  (i.e., time-dependent) and quenched parts of the correlations, respectively. The exponents in (\ref{Correl1}) are
\begin{subequations}
\label{eq:exponents}
\begin{align}
z&= 0.49\pm 0.01,\ \
\zeta=0.77\pm 0.01,\\
\chi&=-0.23\pm 0.01,\ \ \chi'=-0.37\pm 0.02\ ,
\end{align}
\end{subequations}
in {the physical case} $d=2$,
where the error bars correspond to the differences between the
 three aforementioned {DRG} schemes.
 The fact that both $\chi$  and $\chi'$
 are negative implies long-range polar order.

 The existence of a common anisotropy exponent $\zeta$ for the quenched and the annealed parts of the correlation function (\ref{Correl1}) and a simple scaling form for the latter is a highly non-trivial result, which is a direct consequence of the anomalous hydrodynamics of our system.
Indeed, a {\it linearized} treatment of our hydrodynamic model predicts a much more complicated ``double scaling" form, with different anisotropy exponents for the quenched and annealed correlations, and a more complicated form for the annealed part of the correlation function (what we mean by this will become clear below). The latter  is reminiscent of, e.g., simple fluids, whose spatiotemporal correlation functions have a structure that reflects both the dispersionless propagation of sound (which is characterized by a dynamic exponent ($z=1$) and the diffusive nature of viscous damping ($z=2$)).

 {\it Model.---}The hydrodynamic equation of motion (EOM) of
incompressible polar active fluids  moving on a disordered substrate can be constructed based on symmetry considerations alone \cite{toner_prl95,toner_pre98,wensink_pnas12,CLT_Ncomm,CLT_incom2}. We defer its derivation to the {Associated Long Paper (ALP)} \cite{ALP}. The EOM in the moving phase is identical to that studied in \cite{CLT_Ncomm}, except for {the} presence of the quenched noise. Specifically,  keeping only ``relevant" terms, by which we mean terms that can change the long-distance, long-time behavior of the system, the EOM governing  $\bu$  is, in Einstein component notation,
\beqn
\label{eq:u_eom}
\pp_t u_i&=&-\pp_i \Pi-\gamma\pp_xu_i-b\delta_{ix} \pp_x u_x+\mu_\perp \pp_y^2 u_i+f_{_Q}^i+f_{_A}^i
\nonumber\\
&&
+\mu_x \pp_x^2 u_i-\alpha\left(u_x+{u_y^2\over 2v_0}\right)(\delta_{ix}+{u_y\over v_0}\delta_{iy})\,,
\eeqn
 where the indices $i,j$ label the spatial coordinates. In (\ref{eq:u_eom}), the ``pressure" $\Pi$ acts as a Lagrange multiplier to enforce the incompressibility condition: {$\vnab \cdot \bu = 0$}, while $\gamma$ and $b$ are (positive or negative) constants, and $\alpha$ and {$\mu_{x,\perp}$} are positive constants \cite{ALP}.  The quenched and annealed noises $\bff_{_Q}(\br)$ and $\bff_{_A}(\br,t)$ respectively  are zero mean Gaussian white noises with variances:
\begin{subequations}
\label{eq:noise_def}
\begin{align}
\langle f_{_Q}^i(\br)f_{_Q}^j(\br')\rangle&=
2D_{_Q}\delta_{ij}\delta^2(\br-\br')\,,\label{Random_Q}\\
\langle f_{_A}^i(\br,t)f_{_A}^j(\br',t')\rangle&=
2D_{_A}\delta_{ij}\delta^2(\br-\br')\delta(t-t')\, {.}\label{Random_A}
\end{align}
\end{subequations}
Note the time-independence of the
noise $\bff_{_Q}(\br)$; this is what we mean by ``quenched".
{As mentioned earlier,} Eq.~(\ref{eq:u_eom}) differs from the EOM
in \cite{CLT_Ncomm} {only}
through the presence of the quenched noise $\bff_{_Q}(\br)$.

{\it Linear regime.---} {The linearized version of EOM \eqref{eq:u_eom}  is solved in Fourier space by projecting \eqref{eq:u_eom} transverse to the wavevector using the projection operator $P_{li}(\bq)=\delta_{li}-\frac{q_lq_i}{q^2}$. {Autocorrelating these solutions, and using the correlations \eqref{Random_Q} and \eqref{Random_A} for the noises}, we obtain the correlation functions}
\beqn
\label{eq:corre_uu}
\langle u_i(\tilde{\bq})u_i(\tilde{\bq}')\rangle=
 C^i_{_{A}}(\tilde{\bq},\tilde{\bq}')
+C^i_{_{Q}}(\tilde{\bq},\tilde{\bq}')\,,~~~
\eeqn
where {$\tilde{\bf q}\equiv (\omega,{\bf q})$, with $\omega$ being the frequency and ${\bf q}$ the wavevector of the perturbation,}
\begin{subequations}
\label{eq:Ca&Cq}
\begin{align}
\label{2D_linear_A}
C_{_{A}}^{x,y}(\tilde{\bq},\tilde{\bq}')&={q_{y,x}^2\over q^2}{2D_{_A}\delta(\omega+\omega')\delta(\bq+\bq') \over \left[\omega-\left({b q_{y}^2\over q^2}+\gamma\right)q_x\right]^2+\left[{\alpha q_{y}^2\over q^2}+\Gamma(\bq)\right]^2}\,,\\
\label{2D_linear_Q}
C_{_{Q}}^{x,y}(\tilde{\bq},\tilde{\bq}')&={q_{y,x}^2\over q^2}{4\pi D_{_Q}\delta(\omega)\delta(\omega')\delta(\bq+\bq')\over \left({b q_{y}^2\over q^2}+\gamma\right)^2q_x^2+\left[{\alpha q_{y}^2\over q^2}+\Gamma(\bq)\right]^2}\, {,}
\end{align}
\end{subequations}
{where $\Gamma({\bf q}){ \equiv \mu_\perp q_y^2+\mu_x q_x^2}$,}
and the subscripts $A$ and $Q$ denote the annealed and quenched parts, respectively.

The real-time, real-space autocorrelation of $\bu(\br,t)$ obtained via inverse Fourier-transforming \eqref{eq:Ca&Cq}, is
\beqn
\langle \bu(\br,t)\cdot\bu(\mathbf{0},0)\rangle&=&|y|^{-{1\over 3}}{ {\cal C}_{_{Q}}}\left(|x|\over |y|^{2\over 3}\right)\nonumber\\&&+|y|^{-{1\over 2}}{ {\cal C}_{_{A}}}\left({|x-\gamma t|\over |y|^{1\over 2}},{|t|\over |y|}\right)\,,\label{Real_linear_corr}
\eeqn
where  ${\cal C}_{_{A/Q}}$
are scaling functions
displayed in  \cite{ALP}.
Thus the linear theory predicts $\chi=-1/6$,
$\chi'=-1/4$, and $z=1$, and also predicts different anisotropy exponents for the quenched and annealed parts:  $\zeta_{\rm quenched}=2/3$ and  $\zeta_{\rm annealed}=1/2$. Note also that the first argument of the scaling function ${\cal C}_{_{A}}$ in the linear theory involves a ``boosted" $x$ coordinate $x-\gamma t$, in contrast to the non-linear result \eqref{Correl1}, in which the first argument of the annealed scaling function only involves $x$, with no $t$-dependence.
The negativity of the roughness exponents $\chi$ and $\chi'$ implies that the incompressible flock has long-range polar order within the linear theory \cite{Ano_disord}.
 We will show that these exponents  are modified
 in the nonlinear theory,
and, in particular, the two anisotropy exponents $\zeta_{\rm quenched}$ and $\zeta_{\rm annealed}$
become equal.

{This persistence of two-dimensional long-range polar order in a disordered medium is a result of a combination of the effects of incompressibility  and, crucially, active motility, as can be seen from \eqref{eq:Ca&Cq}.
{Indeed, in a 2D equilibrium} divergence-free
magnet, the equal-time correlator of {the magnetization} diverges as $1/\mu_x^2q_x^4$ instead of as $1/\gamma^2q_x^2$ for $q_y=0$ in the presence of quenched disorder.  This divergence is strong enough to destroy \emph{even} quasi-long-range order. This implies that, while incompressible flocks perturbed by \emph{annealed} noise are equivalent to \emph{equilibrium} {divergence-free} magnets \cite{CLT_Ncomm, Kashuba}, they are much more resistant to {quenched} disorder. Due to motility, the quenched disorder is effectively ``annealized" for fluctuations propagating along the ordered direction {$\hat{\bx}$} at a speed $\gamma$: along $q_y=0$,  $C_{_{A}}^y(\bq,\bq')$ and $C_{_{Q}}^y(\bq,\bq')$ both diverge as $\sim 1/q_x^2$  as $\bq\to\mathbf{0}$.}

{\it Nonlinear regime \& DRG analysis.---}
 We now turn to the full EOM of $\bu$ (\ref{eq:u_eom}). Fourier transforming this, and operating on both sides with the transverse projection operator $P_{ly}$, we obtain
\begin{widetext}
\beq
-\ii\omega u_y=P_{yx}\left(\bq\right)\mathcal{F}_{\tilde{\bq}}
\left[-\alpha\left(u_x+{u_y^2\over 2v_0}\right)\right]+\mathcal{F}_{\tilde{\bq}}
\left[-\gamma\pp_xu_y
-\lambda u_y\pp_yu_y-{\alpha\over v_0}
\left(u_x+{u_y^2\over 2v_0}\right)u_y+f_{_A}^y+f_{_Q}^y\right]\,,
\label{2Dy4}
\eeq
\end{widetext}
where we have neglected
terms that are irrelevant
in the dominant regime of wavevector  $q_y\ll q_x$, as we discovered in our treatment of the linear theory. We will
verify {\it a posteriori} that this continues to hold
for the nonlinear theory. The symbol $\mathcal{F}_{\tilde{\bq}}$ represents the $\tilde{\bq}$th Fourier component.

To evaluate the importance of the nonlinear terms in \eqref{2Dy4}, we
rescale time, lengths, and  fields as
\begin{subequations}
\begin{align}
t\to te^{z\ell},~~x\to xe^{\zeta\ell},~~y\to ye^{\ell},~~
u_y\to u_ye^{\ell\chi}\,,
\label{rescale}
\end{align}
\end{subequations}
 implying
 	$u_x\to u_xe^{\left(\chi+\zeta-1\right)\ell}$, and keep the form of the resultant EOM unchanged by absorbing the rescaling factors into the coefficients.
The coefficients of the linear terms $P_{yx}u_x$, $\pp_xu_y$, and the noise strength are rescaled respectively as
\begin{subequations}
\begin{align}
\alpha&\to\alpha e^{\left(z+2\zeta-2\right)\ell},~~
\gamma\to\gamma e^{\left(z-\zeta\right)\ell},~~
\\
D_{_A}&\to D_{_A}e^{\left(z-2\chi-\zeta-1\right)\ell}\,,~~
D_{_Q}\to D_{_Q}e^{\left(2z-2\chi-\zeta-1\right)\ell}\,,
\end{align}
\end{subequations}
and the coefficients of the nonlinear terms $P_{yx}u_y^2$, $u_y\pp_yu_y$, $u_xu_y$, and $u_y^3$ as
\begin{subequations}
\label{pcnonlinear}
\begin{align}
{\alpha\over 2v_0}&\to{\alpha\over 2v_0} e^{\left(z+\chi+\zeta-1\right)\ell},~~
{\alpha\over v_0}\to {\alpha\over v_0}e^{\left(z+\chi+\zeta-1\right)\ell}\,,~~
\\
\lambda&\to\lambda e^{(z+\chi-1)\ell},~~{\alpha\over 2v_0^2}\to {\alpha\over 2v_0^2}e^{\left(z+2\chi\right)\ell}\,.
\end{align}
\end{subequations}
We choose $z$, $\zeta$, and $\chi$ to fix $\alpha$, $\gamma$ and $D_{_Q}$, which control the size of the dominant fluctuations (i.e., those coming from the quenched noise),
which yields
\beq
z={2\over 3}\,,~~~\zeta={2\over 3}\,,~~~\chi=-{1\over 6}\,.
\eeq
As expected, these values of $\zeta$ and $\chi$ are identical to those for the quenched correlations obtained from our linear theory [e.g., see \eqref{Real_linear_corr}].
Note that while the choice of $z$ is formally necessary to restore the EOM to its original form, it does not appear in the correlation function \eqref{Real_linear_corr} because the leading order part of that correlation function is static since the quenched disorder that induces those leading-order fluctuations is.
Its value is different from the $z$ in \eqref{Real_linear_corr}, since the time-dependent part of the fluctuations comes from the subdominant annealed disorder.

Substituting  these values into (\ref{pcnonlinear}), we find the coefficients of the nonlinear terms $P_{yx}u_y^2$, $u_xu_y$, and $u_y^3$ rescale as
\beq
{\alpha\over 2v_0}\to{\alpha\over 2v_0} e^{\ell/6},~~
{\alpha\over v_0}\to {\alpha\over v_0}e^{\ell/6}\,,~~
{\alpha\over 2v_0^2}\to {\alpha\over 2v_0^2}e^{\ell/3}\,,
\label{pcnonlinear1}
\eeq
which clearly
all diverge as $\ell\to\infty$. This implies that these nonlinear terms are \emph{relevant} in the hydrodynamic limit.  In contrast,  $\lambda\to\lambda e^{(z+\chi-1)\ell}$ vanishes, which implies $u_y\pp_yu_y$ is irrelevant and hence can be neglected.

To deal with the relevant nonlinearities,
we will  perform a one-loop DRG calculation using an $\epsilon$-expansion method. Our approach is that of  \cite{forster_prl76,forster_pra77}
 and is explained in detail in the ALP \cite{ALP}.

 To employ the $\epsilon$-expansion method, we rely on analytic continuity to generalize our calculation to dimensions $d>2$. We have two distinct choices for doing this; we can treat the ``soft'' $x$ direction
as one-dimensional and the ``hard" $y$ direction as $d-1$ dimensional or we can take $x$ to be $d-1$ dimensional while treating $y$ as one-dimensional. These lead to two \emph{distinct} $\epsilon$-expansion schemes which we term {the} \emph{hard} and \emph{soft}  continuations, respectively \cite{foot}. Alternatively, we perform an uncontrolled one-loop calculation in exactly $d=2$. The numerical values
 given at the beginning of the Letter for the exponents represent an average of the results from these three schemes.

Crucially, in all three schemes, we make use of an important simplification:
due to the rotation invariance of our hydrodynamic EOM, it is convenient to choose the values of $\chi$ and $\zeta$ so that all three ``$\alpha$"s   appearing in  the EOM (\ref{2Dy4})
remain identical upon the rescaling, i.e.,
\beq
\chi=\zeta-1\,.\label{Chi1}
\eeq
Using the above simplification to eliminate the roughness exponent $\chi$, the DRG recursion relations, to one-loop order, for the {\it hard} continuation are \cite{ALP}:
\begin{subequations}
\label{rr}
\begin{align}
{\dd\ln\alpha\over\dd\ell}&=z+2\zeta-2+\eta_\alpha \,,\label{fl_Alpha}\\
{\dd\ln\gamma\over\dd\ell}
&=z-\zeta+\eta_{\gamma}\,,\label{fl_gamma}\\
{\dd\ln{\mu_x}\over\dd\ell}&=z-2\zeta+\eta_\mu\,,\label{Fl_mu_x_exact}\\
{\dd \ln D_{_Q}\over\dd\ell}&=2z-3\zeta+3-d+\eta_{_Q} \,,\label{fl_D_Q}\\
{\dd \ln D_{_A}\over\dd\ell}&=z-3\zeta+3-d+\eta_{_A}\,,\label{fl_D_A}
\end{align}
\end{subequations}
where the $\eta$'s represent graphical corrections. To   lowest  order in perturbation theory \cite{ALP},
\begin{subequations}
\label{Anoma_Exp1-0}
\begin{align}
\eta_{\alpha}&=-{1\over 27}g\,,~~\eta_{\gamma}={8\over 27}g\,,~~\eta_{_Q}={10\over 27}g\,,~~\eta_{_A}={ 16\over 27}g\,,\label{Anoma_Exp1-1}\\
 \eta_\mu&= g_{\mu}+{2\over 27}g\,,\label{Anoma_Exp1-2}
\end{align}
\end{subequations}
where we've defined
\begin{subequations}
\begin{align}
g&\equiv {S_{d-1}\over (2\pi)^{d-1}}{|}\gamma{|}^{-{7\over 3}}\alpha^{1\over 3}\Lambda^{3d-7\over 3}D_{_Q}\,,\label{g_def}\\
g_{\mu}&= {S_{d-1}\over (2\pi)^{d-1}}\Lambda^{d-3}{|}\gamma{|}^{-1}\mu_x^{-1}D_{_Q}\,,\label{g_mu_def}
\end{align}
\end{subequations}
with $S_{d-1}$ the area of a $d-1$-dimensional unit sphere, and $\Lambda$ the inverse of the short wavelength cut-off in the soft direction.
In writing the recursion relations \eqref{rr}, we have ignored some corrections arising from the annealed noise $D_{_A}$ which prove to be irrelevant (that is, to vanish as $\ell\to\infty$)  \cite{ALP}.

From \eqref{rr}, we can construct  closed recursion relations
for $g$ and $g_{\mu}$:
\begin{subequations}
\begin{align}
{\dd g\over\dd\ell}&={1\over 3}\left(7-3d-g\right)g\,,\label{Flow_g}\\
{\dd g_{\mu}\over\dd\ell}&= \left(3-d-g_{\mu}\right)g_{\mu}\,.\label{Flow_g_mu}
\end{align}
\end{subequations}
For {$d<{7/3}$}, the above flow equations indicate that the generic stable fixed point is at
\beq
g^*=3\epsilon +\cO(\epsilon^2)\,,~~~g_\mu^*=\frac{2}{3}+\epsilon+\cO(\epsilon^2)
\ ,
\label{fix_p_ep_v2}
\eeq
where  $\epsilon=7/3-d$ and
the
$O(\epsilon^2)$ part can only be obtained through higher-order in perturbation theory  calculations.
 For $d=2$, $\epsilon=1/3$, which is extremely small for $\epsilon$ expansions. Therefore, we expect our one-loop DRG results to be quantitatively \emph{very} accurate.

 The fact that both $g$ and $g_\mu$ flow to non-zero stable fixed points also implies two {\it exact} relations between the $\eta$'s.
The definition of $g$ (\ref{g_def}), together with \eqref{rr},
implies
\beq
{\dd\ln{g}\over \dd\ell}={1\over 3}\left(7-3d\right)-{7\over 3}\eta_{\gamma}+{1\over 3}\eta_{\alpha}+\eta_{_Q}\,.\label{Flow_g_1}
\eeq
which leads to the exact relation, valid {to} \emph{all} loop orders,
\beq
7-3d-7\eta_{\gamma}+\eta_{\alpha}+3\eta_{_Q}=0\,,\label{Exact_Relation1}
\eeq
since $\dd \ln g/\dd \ell=0$ at the fixed point.
Similarly, (\ref{g_mu_def}) and \eqref{rr}
{lead to} the second
{\it exact} relation:
\beq
3-d+\eta_{_Q}-\eta_\gamma-\eta_\mu =  0\,.\label{Exact_Relation2}
\eeq

 Note that $\eta_\mu$ does not vanish as $\epsilon$ goes to zero since the upper critical dimension of $g_\mu$ is $3$ instead of $7/3$
(see \eqref{g_mu_def}). Since $\mu_x$ controls the annealed part of the correlation, this implies that this part starts to depart from the linear prediction, as $d$ is decreased, from $d=3$, {in contrast to} the quenched part, which {first does so} at $d=7/3$.
As shown in detail in \cite{ALP}, while {within the linear theory, the} anisotropy exponent for the quenched and annealed parts of the correlator are \emph{distinct}, the annealed anisotropy exponent assumes the same value as the quenched one for all $d\leq7/3${; that is, in precisely the dimension} at which the quenched anisotropy exponent first departs from its linear value.

{\it Scaling exponents.---}We now use the DRG procedure
to calculate the real time-real space correlations $C_{_Q}\left(\br\right)$ and $C_{_A}\left(t,\br\right)$, which represent the quenched and annealed parts of $\langle\bu(\br,t)\cdot\bu(\mathbf{0},0)\rangle$, respectively.
They are related to those of the rescaled system \cite{Nelson_traj} via
\begin{subequations}
\begin{align}
&C_{_Q}\left(\alpha_0,\gamma_0,D_{_{Q0}},\br\right)\nonumber
\\=&\ e^{2\chi\ell}C_{_Q}\left[\alpha(\ell),\gamma(\ell),D_{_Q}(\ell),{|x|\over e^{\zeta\ell}},{|y|\over e^{\ell}}\right]\,,\label{Traje1}
\\
&C_{_A}\left(\alpha_0,\mu_{x0},D_{_{A0}},t,\br\right)\nonumber
\\=&\ e^{2\chi\ell}C_{_A}\left[\alpha(\ell),\mu_x(\ell),D_{_A}(\ell),{|t|\over e^{z\ell}},{|x|\over e^{\zeta\ell}},{|y|\over e^{\ell}}\right]\,,~~~~\label{Traje2}
\end{align}
\end{subequations}
where $\alpha$, $\gamma$, and $D_{_Q}$
control the magnitude of $C_{_Q}\left(\br\right)$, while $\alpha$, {$\mu_x$} and $D_A$ control the magnitude of $C_{_A}\left(\br,t\right)$ and the subscript ``0" denotes the bare values of the parameters.
We choose $\zeta$ and $z$ such that $\alpha$, $\gamma$, $\mu_x$, and $D_{_{Q}}$ are all fixed [i.e., equate the RHS of (\ref{fl_Alpha}-\ref{fl_D_Q}) to 0], which {\it is} possible since $\eta_{\gamma,\alpha,_Q,\mu}$ are constrained by the {\it exact} relations (\ref{Exact_Relation1},\ref{Exact_Relation2}).
Choosing $\ell=\ln\left(\Lambda |y|\right)$ and taking the renormalization factor of $D_{_A}$ to the front, we write the RHS of (\ref{Traje1},\ref{Traje2}) in the form  displayed by (\ref{Correl1}) with
\beq
\cG_{_{Q}}\left(|x|\over |y|^{\zeta}\right)
\equiv\Lambda^{2\chi}C_{_Q}\left[\alpha_0,\gamma_0,D_{_{Q0}},{|x|\over (|y|\Lambda)^{\zeta}},{1\over\Lambda}\right]\,,\label{}
\eeq
{and}
\begin{eqnarray}
&&\cG_{_{A}}\left({|t|\over |y|^z},{|x|\over |y|^{\zeta}}\right)\nonumber\\
&\equiv&	\Lambda^{2\chi'}C_{_A}\left[\alpha_0,\gamma_0,D_{_{A0}},{|t|\over (|y|\Lambda)^z},{|x|\over (|y|\Lambda)^{\zeta}},{1\over\Lambda}\right]\,,
\end{eqnarray}
with the various exponents being
\begin{subequations}
\begin{align}
\zeta=&\ {2+\eta_{\gamma}-\eta_{\alpha}\over 3}={2\over 3}+{1\over 3}\epsilon+O(\epsilon^2)\,,\label{Zeta1}\\
z=&\ {2-2\eta_{\gamma}-\eta_{\alpha}\over 3}={2\over 3}-{5\over 9}\epsilon+O(\epsilon^2)\,,\label{Z1}\\
\chi=&\ {-1+\eta_{\gamma}-\eta_{\alpha}\over 3}=-{1\over 3}+{1\over 3}\epsilon+O(\epsilon^2)\,,\label{Chi1}\\
\chi' =&\ { \eta_{_A}-\eta_{\gamma}-1\over 2}=-{1\over 2}+ {4\over 9}\epsilon+O(\epsilon^2)\,,\label{Chi'}
\end{align}
\end{subequations}
where in the second equalities we have used (\ref{Anoma_Exp1-0}) and the results for $g$ and $g_{\mu}$ to $O(\epsilon)$ given by (\ref{fix_p_ep_v2}). Note that this calculation shows explicitly that, beyond the linear theory, the anisotropy exponents for the quenched and the annealed correlations become equal, and the $x$-dependence of the annealed part of the correlation function involves an ``unboosted" $x$, {\it not} a boosted variable $x-\gamma t$. This confirms our earlier claim.

The analysis for the {\it soft} continuation and the uncontrolled one-loop calculation in precisely $d=2$ are very similar, and give comparable quantitative results \cite{ALP}.
The  exponents quoted earlier are a weighted average of these three results; the quoted error bars more than cover the spread between the three different  calculations.

{\it Summary \& Outlook.---}{We have demonstrated that a combination of active motility and {incompressibility}
	leads to the formation of two-dimensional long-range ordered flocks, even in the presence of random quenched disorder.} {In contrast, neither active motility nor  incompressibility {\it  alone} can overcome quenched disorder, although either can successfully compete with {\it annealed} (i.e., time-dependent) noise \cite{toner_pre12, Kashuba}.}

One experimental realization of incompressibility  in a two-dimensional system  is motile particles moving through a narrow channel filled with an  incompressible fluid \cite{BrottoLauga, Ano_pol} or at high particle densities \cite{CLT_Ncomm, Ano_pol}. Furthermore, some degree of disorder will always be present in all experimental systems, especially biological ones. Therefore, our work should be valuable in interpreting numerous experiments both in artificial and biological active systems.
Further, it demonstrates that confluent cell layers on substrates can move coherently despite the presence of static random impurities.
	
	We also look forward to quantitative tests of our predictions in artificial active systems, for instance, Quincke rotors \cite{Bartolo2021} or vibrated granular systems \cite {Soni}.

\begin{acknowledgments}
L.C. acknowledges support by the National Science
Foundation of China (under Grant No. 11874420). J. T.
 thanks the Max Planck Institute for the Physics of Complex Systems,
Dresden, Germany, for their support through a Martin
Gutzwiller Fellowship during this period. AM was supported by a TALENT fellowship from CY Cergy Paris Universit\'e. L.C. also thanks the MPI-PKS where the early stage of this work was performed for their support.
\end{acknowledgments}


\begin{thebibliography}{16}
\bibitem{foot} We {call} $y$ and $x$ ``hard" and ``soft" directions, respectively, since the {Fourier modes of the velocity with their wavevectors along $y$ have the  smallest fluctuations, while those with their wavevectors along $x$ have the largest.}
\end{thebibliography}
\end{document}